\documentclass[twocolumn,showpacs,amssymb]{revtex4}

\usepackage{graphicx}
\usepackage{dcolumn}
\usepackage{bm}

\def\gtap{\mathrel{ \rlap{\raise 0.511ex \hbox{$>$}}{\lower 0.511ex
   \hbox{$\sim$}}}} 
\def\ltap{\mathrel{ \rlap{\raise 0.511ex
    \hbox{$<$}}{\lower 0.511ex \hbox{$\sim$}}}} 

\def\be{\begin{equation}}
\def\ee{\end{equation}}
\def\bea{\begin{eqnarray}}
\def\eea{\end{eqnarray}}

\def\<{<\!\!}
\def\>{\!\!>}
\def\<{\langle}
\def\>{\rangle}

\begin{document}
\input{epsf}

\title{Correlation between galactic H{\sc i} and 
the Cosmic Microwave Background}

\author{Kate Land$^1$ and An\v{z}e Slosar$^{1,2}$}

\affiliation{$^1$Astrophysics, Denys Wilkinson Building, University of Oxford, 
Keble Road, OX3RH1, Oxford, UK} 
\affiliation{$^2$Faculty of Mathematics and Physics, University of Ljubljana,
  Slovenia}

\begin{abstract}
  We revisit the issue of a correlation between the atomic hydrogen
  gas in our local Galaxy and the Cosmic Microwave Background (CMB), a
  detection of which has been claimed in some
  literature.  We cross-correlate the 21-cm emission of
  Galactic atomic hydrogen as traced by the Leiden/Argentine/Bonn
  Galactic H{\sc i} survey with the 3-year CMB data from the Wilkinson
  Microwave Anisotropy Probe. We consider a number of angular scales, masks, 
and H{\sc i} velocity slices and find no statistically significant
  correlation.
\end{abstract}

\date{\today}
\maketitle

\section{Introduction}

Since the beginning of this millenium significant advances in Cosmology have 
been made, in large part due to the advent of high-quality cosmological
datasets. The full-sky results of the Cosmic Microwave Background
(CMB) fluctuations from NASA's Wilkinson Microwave Anisotropy Probe
(WMAP)~\citep{WMAP_temp} have been particularly spectacular, 
finding excellent agreement with independent cosmological observations and 
the general predictions of our inflationary $\Lambda$CDM 
cosmological model~\cite{WMAP_cosmo}. Specifically, 
the temperature perturbations 
are consistent with a Gaussian random field exhibiting 
a distinct power spectrum with 
peaks due to sound waves in the plasma before the epoch of
recombination.

It is no surprise that the WMAP dataset
has received a lot of attention, with various attempts to find and
characterise anomalies that would signal a primordial departure from
a Gaussian random field or contamination of the data
(e.g.~\cite{Komatsu,Erik07,Land04,Slosar04,Viel03,Liu,
  McEwen04,McEwen06,Erik04,Bernui,Chiang,Huff}). While there are some
indications of possible pathologies in the data, the statistical
significances vary and the general consesus so far is that the data
is not contaminated with foregrounds at a significant level.

However, a recent work has claimed that there are spatial
associations between neutral hydrogen and small-scale CMB
structure observed by WMAP~\citep{versch}. These associations are claimed 
to be significant at scales of order $1^\circ$, which also 
corresponds to the scale of the first acoustic peak in the 
CMB power spectrum. This peak is an important ``standard ruler'' when it 
comes to parameter fitting and thus any such contamination could 
have serious consequences for the inferred power spectrum and
cosmological parameter values.
However, the statistical significance of these correlations are currently 
unclear. Visual inspection alone is insufficient as the human brain is 
particularly susceptible to
discerning structure in random fields.
After all, one can easily be convinced that Stephen 
Hawking's initials are imprinted in the WMAP \emph{Internal Linear 
Combination} (ILC) map at $(l,b)\sim(60^\circ,10^\circ)$.

In this paper we extend the work of~\cite{versch} and perform a
rigorous examination of the correlation between the third-year 
single-frequency WMAP
maps and tracers of H{\sc i} structure from the Leiden/Argentine/Bonn
Galactic H{\sc i} Survey (LAB) experiment~\citep{LAB}. There is no
established mechanism by which the Galactic neutral hydrogen could 
affect the WMAP data, besides the fact that it spatially correlates with other
contaminants. Thus, a statistically significant correlation
would indicate that either there is a previously unknown emission
process taking place or, more likely, that the CMB data has not been
adequately cleaned or masked.

In this brief report we perform a series of tests, employing various 
sky-cuts and probing a range of angular scales. We find no 
statistically significant correlations. In Section~\ref{sec:data-method}
we present the datasets that we use, outline our
cross-correlation method, and outline the process by which we will assess 
statistical significance. In Section~\ref{sec:results} we present our
results, while the last Section~\ref{sec:conclusions} concludes the paper.


\section{Data and method}
\label{sec:data-method}

\begin{figure}
\centerline{\includegraphics[angle=90,width=8.0cm]{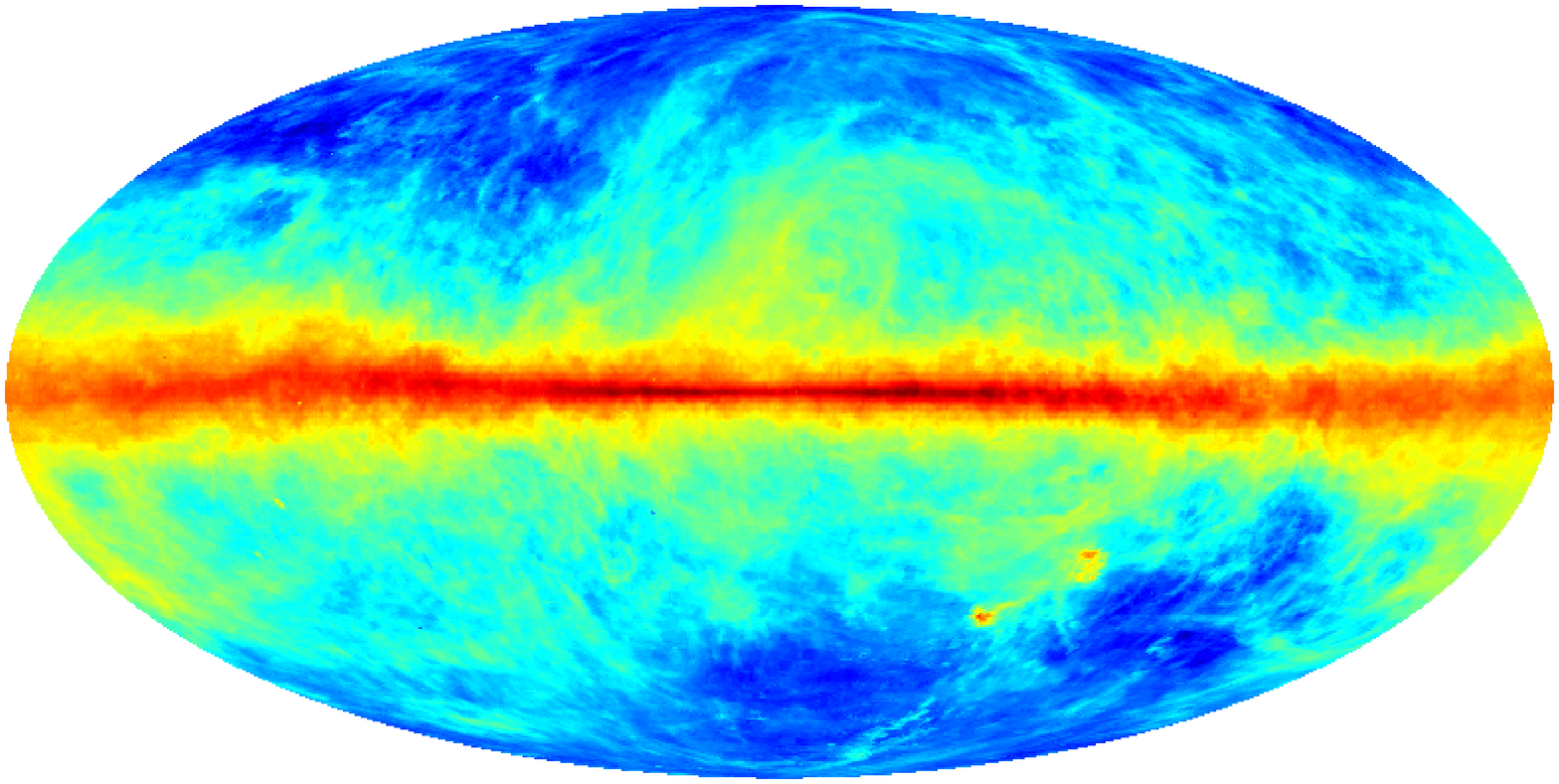}}
\centerline{\includegraphics[angle=90,width=8.0cm]{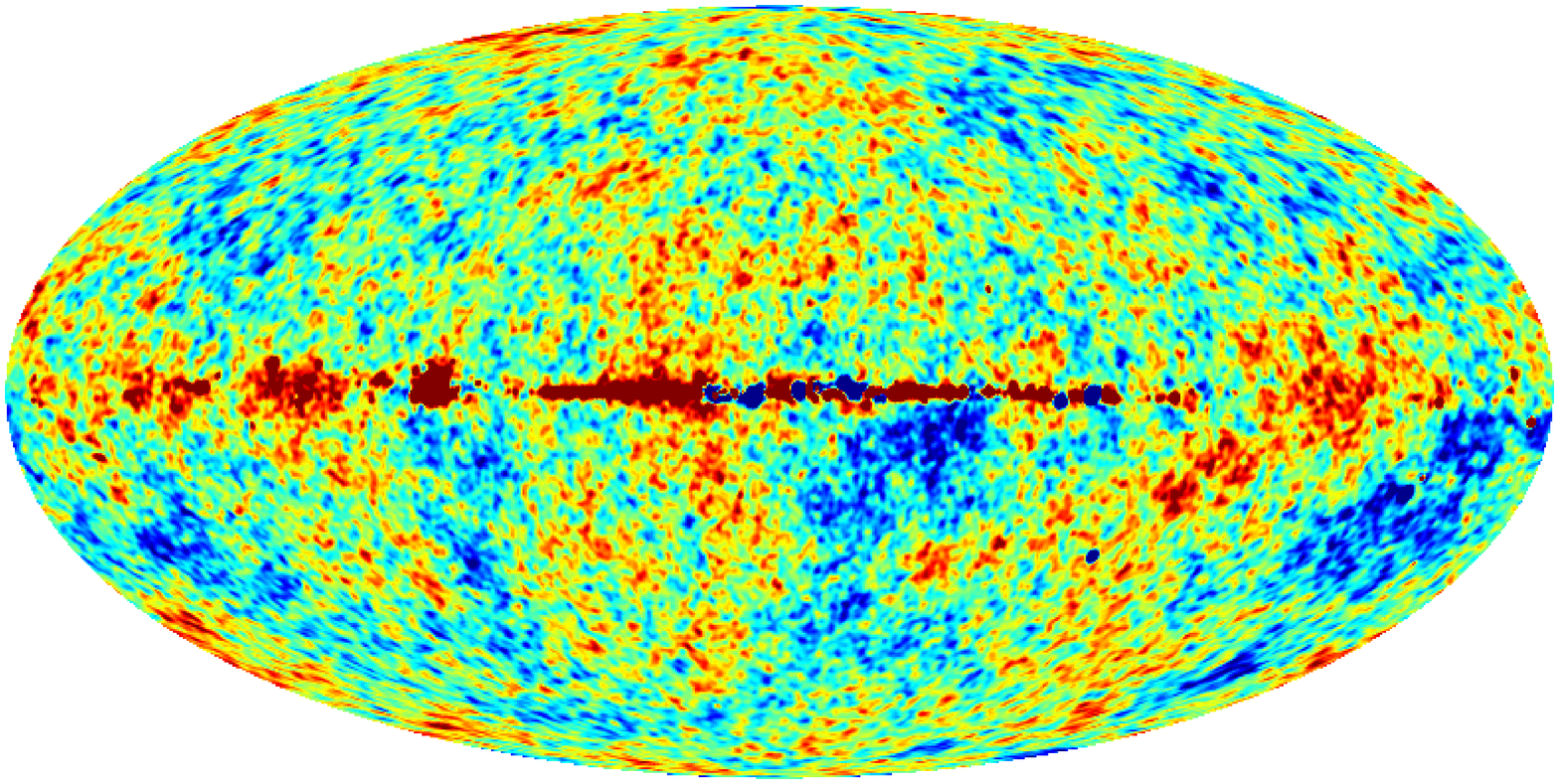}}
\caption{The Full velocity LAB map (above) and the foreground reduced
  V-band WMAP map (below), in galactic coordinates and Mollweide
  projection. The scales are logarithmic in column density and linear
  in temperature fluctuation. }
\label{fig:FV}\end{figure}

As a tracer of the neutral hydrogen we use data from the LAB
survey~\cite{LAB}. The data is distributed in a 3-dimensional cube
spanning the full-sky with half-degree steps in galactic longitude and
latitude. The third component traces the line of sight velocity $v$
spanning the interval $-$450 km/s to $+$400 km/s. We have resampled
the survey into a HEALPix~\citep{healpix} map with $N_{\rm side}=512$.
We tried several schemes of resampling and found that the results are
independent of the exact scheme used. Results reported in this work
correspond to the nearest-neighbour resampling. We have prepared the
following H{\sc i} maps:
\begin{itemize}
\item \emph{Full velocity} (FV) ;
 An integral of all hydrogen data along the line of sight.
\item \emph{High velocity} (HV) ;
 Line of sight integral of $v < -$100 km/s.
\item \emph{Intermediate velocity} (IV) ;
 Line of sight integral of $-$100 km/s $ < v < -$30 km/s.
\item \emph{Low velocity} (LV) ;
 Line of sight integral of $-$30 km/s $< v < +$30 km/s.
\item \emph{Velocity cuts} ;
 Various line of sight integrals over a 10km/s portion of the 
velocity range $-$450 km/s $ < v < +$400 km/s. Resulting in 85 maps, 
herein named according to their mean velocity, 
i.e. $-$445, $-$435, ..., $+$395 km/s.
\end{itemize}
The HV, IV, LV maps are defined to match those used by~\cite{versch}. 
In Figure~\ref{fig:FV} we show the FV map.

For the CMB data we use the three-year WMAP foreground reduced 
temperature maps for the Q, V, and W frequency bands
~\footnote{Available at http://lambda.gsfc.nasa.gov/}. 
These were produced (by the WMAP collaboration) 
by subtracting synchrotron, free-free, and dust emission 
templates from the single year ``unreduced'' maps. Full three-year maps
consist of single year maps coadded with inverse noise weighted 
coefficients, and the single frequency maps consist of the 
relevant differencing assembly data similarly coadded. In Figure~\ref{fig:FV} 
we show the V-band map (smoothed with a beam of $1^\circ$ FWHM).

The CMB data remains contaminated by galactic emission and a mask of 
the galactic plane is necessary. We employ three different masks.
\begin{itemize}
\item \emph{Kp2} ; The standard CMB mask, produced by the WMAP team (a 
15.3\% cut)
\item \emph{Kp0} ; A more aggressive intensity mask (a 23.5\% cut)
\item \emph{Rectangle} ; A mask exposing just the window 
$60^\circ < l < 180^\circ$, $30^\circ < b <70^\circ$, as used 
in~\cite{versch}.
\end{itemize}

We perform the cross-correlation in harmonic space, over the range
$\ell=2-200$, with the upper limit reflecting the resolution of LAB.
The cross-correlation of the angular multipole $\ell$ is calculated
according to \be X_\ell =\frac{1}{(2\ell+1)}\sum_{m=-\ell}^{\ell} a^{\rm CMB}_{\ell
  m} a^{* {\rm HI}}_{\ell m},  \ee where the $a_{\ell m}$ are the 
spherical harmonic coefficients of the respective maps.
The so-defined quantity is
rotationally invariant and real.  Assuming the datasets are Gaussian
and uncorrelated the quantity $X_\ell$ will be normally distributed in the 
limit of large $\ell$ (by the central limit theorem), with corresponding 
standard deviation $\Delta X_\ell$ given by \be (\Delta X_\ell)^2
\sim \frac{1}{f_{\rm sky}\:(2\ell+1)} C^{\rm CMB}_\ell C^{\rm
    HI}_\ell,\label{cosmicvar}\ee where $f_{\rm sky}$ is the fraction of sky being
observed.  However, the LAB dataset is not expected to be Gaussian,
and the presence of the mask correlates the maps somewhat. Therefore
we determine the variance $\Delta X_\ell$ (and mean $\bar{X_\ell}$) 
by performing Monte-Carlo
cross-correlations of the LAB maps with 2000 simulated WMAP maps, 
and we determine these statistics separately 
for all combinations of WMAP band, 
mask and LAB map.
The WMAP simulations are created assuming the third-year WMAP only best-fit
theoretical power spectrum, smoothed with the 
beam profiles of the relevant frequency band, and with noise 
added according to the $N_{\rm  obs}$ field in the individual maps. 
We assume the noise between pixels is uncorrelated; an approximation that 
holds very well for the WMAP temperature maps~\citep{WMAP_temp}.

The effective $\chi^2$ is then calculated, \be \chi^2 =
\sum_{\ell=\ell_{\rm min}}^{\ell_{\rm max}} \left(
  \frac{X_\ell-\bar{X_\ell}}{\Delta X_\ell} \right)^2.  \ee Assuming
independent multipoles this variable should be approximately 
$\chi^2$ distributed with $(\ell_{\rm max}-\ell_{\rm min}+1)$ degrees
of freedom. However, to be more accurate we compare our resulting
$\chi^2$ values to those from the simulations (for the same frequency
band, LAB map, mask and $\ell$-range). Significance of the departure
from a null-correlation is thus assessed through the percentage of
simulations with lower $\chi^2$ values.

As a simple test, we have performed analysis with unmasked data and
we observe a very large correlation that systematically varies with QVW
band, as expected due to the neutral hydrogen tracing known contaminants 
such as dust, free-free, and synchrotron emission.


\section{Results}
\label{sec:results}

\begin{figure}
\centerline{\includegraphics[width=8.0cm]{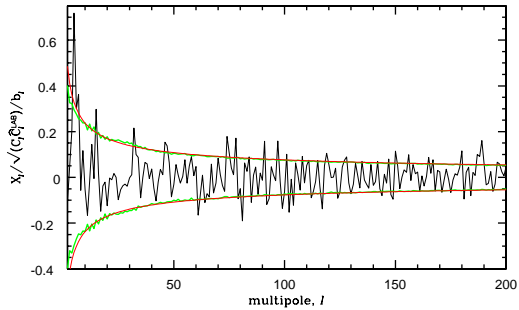}}
\caption{The cross-correlation of the Full velocity 
LAB survey with WMAP V-band, outside the KP2 mask, divided by the power 
spectra and V-band beam. Also plotted 
are the 1-sigma contours calculated from 5000 simulations (green thick) and 
from Eqn(\ref{cosmicvar}) (red thin). The equivalent 
plot for the Q and W-bands are essentially identical to this one.
} 
\label{fig:X}\end{figure}

We consider 89 LAB maps in total, and 3 different masks. We
further focus our attention on 3 different $\ell$-ranges: 2-200 to
cover all the scales, 2-20 for large scales only, and 160-200 for near
degree scales. We compute the $\chi^2$ values for all possible 
combinations of the 3 masks, 3 WMAP bands, 89 LAB maps, and 3 $\ell$-ranges 
(using $\bar{X_\ell},{\Delta X_\ell}$ values determined for the same 
mask-band-map combination). 
In Figure~\ref{fig:X} we plot an example cross-correlation result 
$X_\ell$, with 1-sigma contours returned from simulations and from 
Eqn (\ref{cosmicvar}). 

In Table~\ref{tab:1} we consider how many of our
$\chi^2$ values are higher than 95\% of those from simulations, 
and we find a total of 112 cross-correlation results high at this 
level. By definition, from a total of 
$89 \times 3 \times 3 \times 3 = 2403$ different $\chi^2$ numbers, 
we expect approximately 120 of our results to be 
significant at this level. Our results are therefore consistent with the 
null hypothesis of no correlation between the WMAP and LAB maps.

\begin{figure}
\centerline{\includegraphics[width=8.0cm]{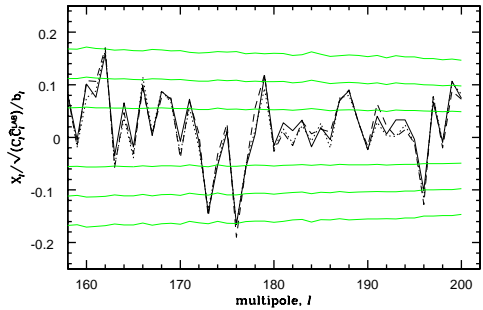}}
\caption{The cross-correlation of the 215 H{\sc i} map with the WMAP data, 
outside the Kp2 mask, over scales $\ell=160-200$ where the $\chi^2$ is found 
to be high at the 99.93\% level. To compare results from the different bands 
[Q(solid), V(dotted, W(dashed)] we have divided out by their repective 
beams. 1,2 and 3-sigma results are also displayed (green solid).
} 
\label{fig:210}\end{figure}

The number of significant results is binomially distributed, with the 
scatter depending on the number of \emph{independent} 
results. Our 2403 results are
definitely not independent: the Q,V,W results are almost identical (as
can be deduced from the Table~\ref{tab:1} and Figure~\ref{fig:210}), 
the $\ell$-ranges overlap, the masks
are not exclusive, and the neighbouring velocity slices are highly
correlated.  This increases the expected spread relative to 
the mean, $(\sigma/\mu)^2\sim(1-p)/(Np)$, and thus from 
further Monte Carlo simulations 
we can determine the effective number of 
independent results that we have. From fitting a binomial distribution 
to a histogram of the number of significant 
(at the 95\% level) $\chi^2$ results that 500 new simulations find, 
we observe an effective $N$ of $\sim 74$. We find a 
consistent result by fitting to the equivalent 5\% confidence level 
results. We find $(\mu,\sigma) \sim (126,66)$, which provides 
further evidence that a total of 112 
signficant values is consistent with no correlation.

  \begin{table}
    \centering
    \begin{tabular}{|c|ccc|}
\hline
      & KP2     & KP0         & RECT \\
\hline
2-200 &  3,5,4 &     3,6,1       &  2,2,1 \\
2-20  &  1,1,1    &  1,1,1   &     0,0,0\\
160-200 & 10,15,8 &  10,14,4  &  5,9,4 \\ 
\hline
    \end{tabular}
    \caption{The number of cross-correlations with 
      $\chi^2$ value that is significant at the 95\% level 
      when compared to simulations. Numbers in each cell correspond 
to the Q,V and W-bands respectively.}\label{tab:1}
  \end{table}

We now turn to extreme values. The highest $\chi^2$ value was obtained
for the correlation of W-band CMB with a H{\sc i} slice at 215 km/s (on 
degree scales, with the Kp2 mask), with only
0.07\% of the W-band simulations finding a higher such 
$\chi^2$ (we did a special
run with $10^4$ simulated maps for this case). How unlikely
is it to have such an extreme value given our sample size? 
If one draws $N$ numbers between $0$ and $1$, the probability
that they will all be less than $x$ is $p=x^N$. For $x=0.9993$
and $N=2403$, $p=0.19$, and therefore it appears entirely typical to get 
such an extreme value for this sample size if the results are independent.
An improved calculation uses the effective number of independent
results, determined to be $\sim 74$, for which we find $p=0.95$ -
indicating that there is a $5$\% probability of obtaining such an
extreme outlier. This result is very conservative, as in reality we
have had more than $74$ numbers from which to select our extreme
value, benefitting from any extra variation.  By considering
limiting cases we see that the probability of our extrema lies
somewhere between 5\% and 81\%, and thus not unusal at any significant
level (ie. above 3$\sigma$).

We note that in contrast with~\cite{versch} we do not observe any 
systematic correlation between the HV, LV, IV, or FV H{\sc i} maps 
and the CMB. Moreover, we do not observe a significant 
correlation between H{\sc i} and the CMB in the area of the sky defined 
by the Rectangular mask, as just 2 of the 89 maps find some correlation 
above the 99\% level, 
and for only one of the $\ell$-ranges - consistent with a chance occurence.

We do, however, observe some correlation on degree scales for a number of 
velocity slices, as seen in the third row of Table~\ref{tab:1}. 
The 15 maps that demonstrate a Kp2 masked V-band degree scale
correlation above the 95\% level 
correspond to mean velocities of -405,
-345, -325, -315,-205, 105\ldots155, 215\ldots235, 345 km/s. Since this
is an {\it a-posteriori} observation and adjacent LAB slices are
highly correlated, it is hard to ascertain the statistical
significance of these. Nevertheless, we track the correlation of
105\ldots155 km/s and 215\ldots235 km/s maps to the same extended feature at
$(l,b)\sim(-50^\circ,-45^\circ)$ (none of the remaining 15 maps are 
significant above the 99\% level). The fact that the correlation is 
due to just one feature again indicates how correlated the slices are, and 
effectively reduces the number of correlations in this case to just one.
In Figure~\ref{fig:210} we examine the correlation of 215 km/s with 
the different 
frequency bands to ascertain if the signal is due to some kind of 
contamination from the presence of this extended H{\sc i} feature. 
However, the signal 
remains very consistent between the CMB bands indicating that there 
is no obvious foreground contamination on these scales. The 
correlation also appears quite random in nature, with no clear trend 
apparent in Figure~\ref{fig:210}.

As discussed, 
some chance correlation between the LAB and WMAP data is expected and we 
do not observe any more than usual. Further, demonstrated by the 
low effective number of independent results (74), the 
velocity slices are highly correlated. Therefore correlations 
with one slice will inevitable lead to correlations with a set of 
neighboring velocity slices. Supported by 
the random nature of the signal in Figure~\ref{fig:210} we conclude 
that this particular set of significant $\chi^2$ values does not 
demonstrate a deviation 
from the norm, but rather it is a manifestation of one of the inevitable 
chance correlations.

\section{Conclusions}
\label{sec:conclusions}

Correlations between Galactic H{\sc i} and the CMB are not expected.
Thus looking for such correlations can be a useful blind test of 
potential systematics, such as foreground contamination of the 
CMB maps. However, there are two important caveats.

Firstly, correlations will inevitably appear from random fluctuations
and one must not use {\it a-posteriori} statistics to claim detections. Even
if one is forced to do so, the required significance bar should be
significantly higher than in the case of expected results. 
Secondly, correlations by eye are very misleading and quantitative 
methods must be employed. Today's Monte-Carlo methods allow
for easy assessment of significance.

We tested for correlation between the third-year WMAP CMB maps and LAB data 
of Galactic H{\sc i}. We considered three different masks, CMB frequency 
bands, angular scales and 89 different H{\sc i} velocity slices. 
We do not find any convincing evidence for a correlation. 
The lack of correlation demonstrates how impressively 
clean the WMAP CMB maps are, outside of the masked regions.


\section*{Acknowledgements}
We thank the anonymous referee for their 
helpful suggestions, and taming the more cavalier parts of this report. 
 KRL is funded by a
Glasstone research fellowship and Christ Church college, AS by Oxford
Astrophysics.


\bibliography{lab}

\end{document}